# Incremental Control System Design and Flight Tests of a Micro-Coaxial Rotor UAV

Zhenchang Liu,[1]
*Science and Technology on Complex System Control and Intelligent Agent Cooperation Laboratory, Beijing, 100074, People's Republic of China*

Yufei Zhang.[2]  Haixin Chen[4]
*Tsinghua University, Beijing, 100084, People's Republic of China*

Zhiduo Zhang[3]
*Beijing Institute of Technology, Beijing, 100081, People's Republic of China*

In this paper, the incremental nonlinear dynamic inversion (INDI) method is applied to the control system design of coaxial rotor UAVs. The aerodynamic uncertainty and anti-disturbance problems are solved in the control system design. The designed controller gives the UAV excellent flight performance and control robustness. An incremental gain method (IGM) is proposed for the delay problem of the state derivative. This method has the advantages of less calculation load and simple parameter adjustment, which provides excellent convenience for applying INDI controllers to coaxial rotor UAVs. The principle of IGM is demonstrated by the stability analysis method of the discrete system, and the parameter selection strategy of the IGM is analyzed in detail by simulation. The advantages of the controller design method are verified by comparative flight tests of nonlinear dynamic inversion (NDI) and INDI. The experimental results show that the average trajectory tracking error of INDI is only 58.3% of that of NDI under the same wind speed. When the angular acceleration delay is less than 0.06 s, IGM can easily keep the system stable. In addition, INDI has better robustness under obvious model errors and strong input disturbance.

## 1. Introduction

Based on developments in avionics and control technology, the dual-motor differential control scheme dramatically simplifies the structure and weight of the coaxial rotor system. This change enables the coaxial rotor system to be used in micro-UAVs [1]. Compared with multi-rotors, coaxial rotors have better thrust efficiency and hovering performance under the same rotor area [2,3]. Moreover, the coaxial rotor UAV can be folded into a cylindrical shape, which is convenient for personal carrying and cluster transportation. The above advantages give coaxial rotor UAVs essential application value in the civil and military fields. It can be used as a portable UAV for aerial photography and surveillance, a gun-launched UAV for barrel launch [4-6], or a subsystem for cluster distribution by the carrier UAV. In addition, coaxial rotor UAVs have the mobility advantages of vertical take-off and landing, hovering, low-speed flight, etc., which have defeated airships and fixed-wing aircrafts and become the preferred scheme for Mars exploration [7-9]. The "Ingenuity" helicopter with a coaxial rotor system landed on Mars in February 2021 and carried out related missions. Compared with the traditional Mars rover, the coaxial rotor UAV greatly expands the detection range and efficiency on the surface of Mars.

---

[1]  Senior Engineer, School of Aerospace Engineering.

[2]  Associate Professor, School of Aerospace Engineering.

[3]  Master Student, School of Aerospace Engineering.

[4]  Professor, School of Aerospace Engineering, Corresponding author, chenhaixin@tsinghua.edu.cn

Coaxial rotor UAVs have apparent coupling and nonlinear control characteristics. There are complex aerodynamic interactions between the upper and lower rotors, and the rotor wake has an interference effect on the fuselage [10,11]. The above factors cause the aerodynamic force and moment of UAVs to have significant uncertainty. It is necessary to model the aerodynamic force and moment of the UAV in the control system design. The modeling error caused by aerodynamic uncertainty affects the control performance. In coaxial rotor control system research, Mokhtari et al. used an extended state observer [12] and a finite-time convergence observer [13] to solve aerodynamic uncertainty and disturbance problems. Song et al. [14] studied an adaptive fault-tolerant control method for coaxial rotor actuator failure based on the backstepping control method. However, the above research is at the simulation level, and the control method is not verified by actual flight tests. Wei et al. [15-16] combined the sliding mode and PID methods to design a coaxial rotor controller and verified the effectiveness and robustness of the controller by flight tests. However, in the above research, only the control performance of coaxial rotors is verified, and no experiments are carried out to test the vehicle performance under external disturbance.

The incremental nonlinear dynamic inversion (INDI) method eliminates the model's aerodynamic uncertainties and disturbances by solving the control incrementally. This method has the advantages of low dependence on model accuracy, good anti-disturbance, and good nonlinear control characteristics [17]. In recent years, the INDI method has been widely used in UAVs [18-27]. In Refs. [23-26], stable INDI controllers were designed based on the time scale separation method. The robustness of INDI to aerodynamic uncertainty and external disturbance was verified through flight tests. Wang et al. [28] used Lyapunov-based methods to prove the stability of any system using INDI. The robustness of INDI was demonstrated from two aspects: theoretical analysis and numerical simulation. In the application of INDI, it is necessary to obtain the derivative of the flight state. The filtering function of the differentiator causes delay in calculating the state derivation, which affects the stability of the control system. Refs. [19,22-26] adopted the active delay method to make the time between the control variable and its increment consistent to avoid system divergence. Lu et al. [29] proposed a method to improve the delay stability of a single input single output (SISO) system using incremental gain and proved this method in detail under the assumption of the continuous system. In Ref [30-32], the Kalman filter, neural network, and other prediction methods were used to obtain the state derivative without delay. However, the above methods have the limitations of complex calculation or model dependence. Sieberling et al. [33] proposed a polynomial prediction method to obtain the no delay state derivative. This method greatly simplifies the calculation of prediction filtering, but it needs additional reference information of the state variables.

According to the existing research and problems of coaxial rotor UAVs, the main contributions of this paper are as follows.

(1) For the first time, the INDI is applied to the control system of the coaxial rotor UAVs, and the incremental gain method (IGM) is developed to solve the problem of system instability caused by angular acceleration delay. The coaxial rotor UAVs exhibits excellent performance in dealing with model uncertainty and external disturbance through the INDI controller.

(2) The improvement principle of IGM on the derivative delay stability of the SISO linear system is proved theoretically using the discrete system stability analysis method. The root locus of incremental gain is acquired, the Bode and Nyquist diagram corresponding to different incremental gains are analyzed in the z domain.

(3) The anti-delay ability, model error correction ability, and parameter adaptability of IGM in the multiple input multiple output (MIMO) nonlinear system are studied through mathematical simulation. The parameter selection strategy of IGM is summarized.

Coaxial rotor UAV hardware in the loop (HITL) simulation and flight tests are conducted to verify all the above contributions.

## 2. Coaxial Rotor Model

The definition of a coaxial rotor UAV and its reference frames are shown in Fig. 1, in which $Ox_b y_b z_b$ is the body reference frame and $Ox_g y_g z_g$ is the ground reference frame. The thrust direction of the coaxial rotors is fixed and opposite to the $Oz_b$ axis. The yaw motion of the UAV is controlled by the speed difference between the upper and lower rotors, and the pitch and roll motions are controlled by the swashplate tilt angle.

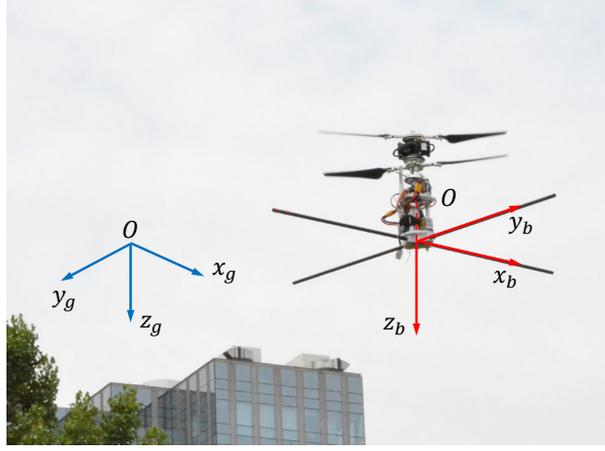

**Fig. 1 Reference frames of coaxial rotor UAV**

### 2.1 6-DOF kinematics and dynamics model

The UAV is regarded as a rigid body for the convenience of analysis. A 6-DOF model is established in the body reference frame, as shown in Eq. (1).

$$\begin{cases} \dot{\mathbf{x}}_0 = \mathbf{L}_{gb}\mathbf{x}_1 \\ m\dot{\mathbf{x}}_1 + m\mathbf{x}_3 \times \mathbf{x}_1 = \mathbf{F} \\ \dot{\mathbf{x}}_2 = \mathbf{R}_{(\phi,\theta)}\mathbf{x}_3 \\ \mathbf{I}\dot{\mathbf{x}}_3 + \mathbf{x}_3 \times \mathbf{I}\mathbf{x}_3 = \mathbf{M} \end{cases} \quad (1)$$

In Eq. (1), $\mathbf{x}_0 = \begin{bmatrix} x & y & z \end{bmatrix}^T$ represents the position information of UAV in $Ox_g y_g z_g$; $\mathbf{x}_1 = \begin{bmatrix} u & v & w \end{bmatrix}^T$ represents the velocity information in $Ox_b y_b z_b$; $\mathbf{x}_2 = \begin{bmatrix} \phi & \theta & \psi \end{bmatrix}^T$ represents the attitude angle information; $\mathbf{x}_3 = \begin{bmatrix} p & q & r \end{bmatrix}^T$ represents the angular velocity information; $m$ is the vehicle mass; $\mathbf{I} \in \mathbb{R}^{3 \times 3}$ is the moment of inertia matrix; $\mathbf{F} \in \mathbb{R}^3$ is the resultant force of the UAV in the body reference frame; $\mathbf{M} \in \mathbb{R}^3$ is the resultant moment; and $\mathbf{R}_{(\phi,\theta)} \in \mathbb{R}^{3 \times 3}$ can be expressed as

$$\mathbf{R}_{(\phi,\theta)} = \begin{bmatrix} 1 & \sin\phi\tan\theta & \cos\phi\tan\theta \\ 0 & \cos\phi & -\sin\phi \\ 0 & \dfrac{\sin\phi}{\cos\theta} & \dfrac{\cos\phi}{\cos\theta} \end{bmatrix}. \qquad (2)$$

$\mathbf{L}_{gb} \in \mathbb{R}^{3\times 3}$ is a rotation matrix from the body reference frame to the ground reference frame, where $\mathbf{L}_{gb}$ is given as follows:

$$\mathbf{L}_{gb} = \begin{bmatrix} \cos\theta\cos\psi & \sin\theta\sin\phi\cos\psi - \cos\phi\sin\psi & \sin\theta\cos\phi\cos\psi + \sin\phi\sin\psi \\ \cos\theta\sin\psi & \sin\theta\sin\phi\sin\psi + \cos\phi\cos\psi & \sin\theta\cos\phi\sin\psi - \sin\phi\cos\psi \\ -\sin\theta & \sin\phi\cos\theta & \cos\phi\cos\theta \end{bmatrix}. \qquad (3)$$

**2.2 Forces and moments acting on the vehicle**

The resultant force of the UAV can be expressed as

$$\mathbf{F} = \mathbf{A} + \mathbf{L}_{gb}^{\mathrm{T}}\mathbf{G} + \mathbf{T} + \mathbf{D}. \qquad (4)$$

In Eq. (4), $\mathbf{A}$ is the aerodynamic force acting on the fuselage in $Ox_b y_b z_b$, $\mathbf{G}$ is gravity, $\mathbf{T}$ is the rotor thrust in $Ox_b y_b z_b$, and $\mathbf{D}$ is the external interference force. Because $\mathbf{A}$'s size is influenced by the rotor wake velocity, body motion speed, and natural wind speed, it is difficult to estimate accurately. Therefore, aerodynamic uncertainty is introduced into the dynamic model of the coaxial rotor UAVs. The installation position of the swashplate and propeller is shown in Fig. 2. Unlike Refs. [12-14], the propeller is installed at the swashplate's AB position, and the thrust direction is changed by the tilt angle of the propeller plane. The propeller of the coaxial rotor UAV studied in this paper is installed at the CD position, and the propeller plane is always perpendicular to the $Oz_b$ axis. The attitude of the UAV is controlled by the cyclic pitch of the propeller.

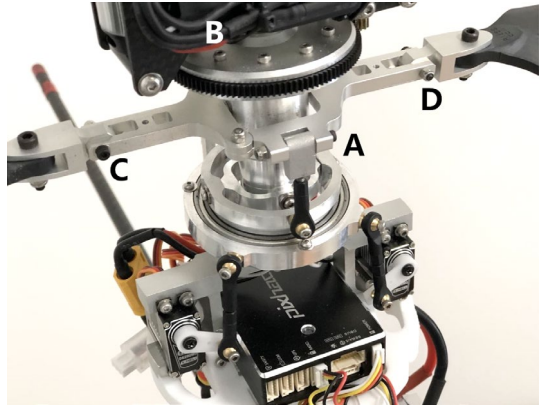

Fig. 2 Swashplate and propeller position

Therefore, **T** can be expressed as

$$\mathbf{T} = \begin{bmatrix} T_x \\ T_y \\ T_z \end{bmatrix} = \begin{bmatrix} 0 \\ 0 \\ \sigma\left(\lambda_1 \omega_1^2 + \lambda_2(\delta_{cx},\delta_{cy})\omega_2^2\right) \end{bmatrix}. \quad (5)$$

In Eq. (5), $\delta_{cx}$ and $\delta_{cy}$ are the tilt angles of the lower rotor along axes $Ox_b$ and $Oy_b$, respectively, and define clockwise rotation as positive. $\lambda_1$ and $\lambda_2$ are the lift coefficients of the upper and lower rotors, respectively, and $\lambda_2$ is a function of $\delta_{cx}$ and $\delta_{cy}$. $\sigma$ is the thrust loss coefficient of the coaxial rotors in the $Oz_b$ direction.

The resultant moment of UAV can be expressed as follows:

$$\mathbf{M} = \mathbf{M}_A + \mathbf{M}_T + \mathbf{M}_D. \quad (6)$$

In Eq. (6), $\mathbf{M}_A$ is the aerodynamic moment, $\mathbf{M}_T$ is the rotor control moment, and $\mathbf{M}_D$ is the external interference moment. $\mathbf{M}_A$ introduces aerodynamic uncertainty to the dynamic model of the coaxial rotor UAVs. The upper and lower rotors turn in opposite directions and rotate at the same speed in the stable hovering state. The angular momentum produced by them counteracts each other, so the control moment does not have a 90-degree lead angle. According to the structural characteristics of the UAV and the principle of cyclic pitch control, $m_T$ and $l_T$ are functions of $\delta_{cx}$ and $\delta_{cy}$, respectively. Because the attitude control of the coaxial rotor UAV is sensitive to the tilt angle of the swashplate, the change in $\delta_{cx}$ and $\delta_{cy}$ is minimal during flight. Therefore, it can be approximated that $m_T$ and $l_T$ are linear with $\delta_{cx}$ and $\delta_{cy}$, respectively. Therefore, $\mathbf{M}_T$ can be expressed as

$$\mathbf{M}_T = \begin{bmatrix} l_T \\ m_T \\ n_T \end{bmatrix} = \begin{bmatrix} k_l \delta_{cy} \\ k_m \delta_{cx} \\ \varsigma_1 \omega_1^2 - \varsigma_2 \omega_2^2 \end{bmatrix}. \tag{7}$$

In Eq. (7), $k_l$ and $k_m$ are the coefficients of roll and pitch control moments, $\omega_1$ and $\omega_2$ are the rotational velocities of the upper and lower rotors, and $\varsigma_1$ and $\varsigma_2$ are the yaw moment coefficients of the upper and lower rotors, respectively.

## 3. Coaxial Rotor Controller Design

According to the time scale separation method, a hierarchical controller is designed for the coaxial rotor UAV. Based on Eq. (1), the controller is divided into four loops: translational kinematics, translational dynamics, rotational kinematics, and rotational dynamics. The structure of the control system is shown in Fig. 3.

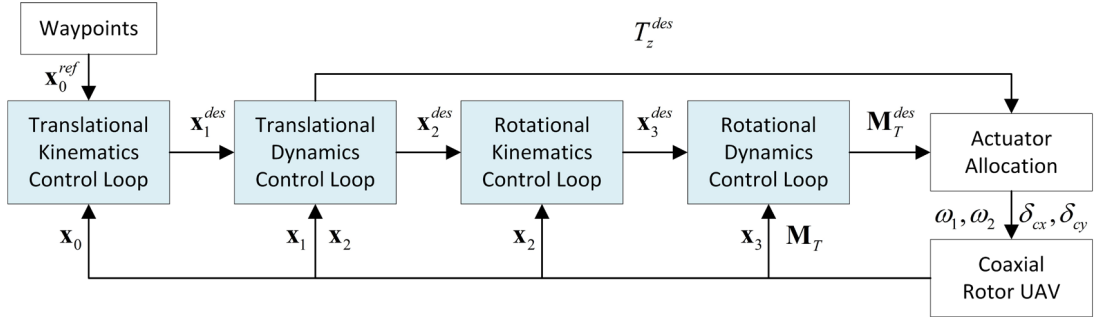

**Fig. 3 Structure of the control system**

In Fig. 3, the superscript "des" represents the desired control variable calculated by the controller, and the superscript "ref" means the reference control variable of the external input. Based on the framework shown in Fig. 3, the NDI and INDI are used to design the controller.

### 3.1 NDI controller design

According to Eq. (1), the NDI controller of the UAV's kinematics loops can be written as

$$\mathbf{x}_1^{des} = \mathbf{L}_{gb}^{-1} \dot{\mathbf{x}}_{0n}, \tag{8}$$

$$\mathbf{x}_3^{des} = \mathbf{R}_{(\phi,\theta)}^{-1} \dot{\mathbf{x}}_{2n}. \tag{9}$$

In Eqs. (8) and (9), $\dot{\mathbf{x}}_{0n} = \mathbf{k}_0 \left( \mathbf{x}_0^{ref} - \mathbf{x}_0 \right)$ and $\dot{\mathbf{x}}_{2n} = \mathbf{k}_2 \left( \mathbf{x}_2^{des} - \mathbf{x}_2 \right)$, where $\mathbf{k}_0, \mathbf{k}_2 \in \mathbb{R}^3$ are the control gains.

In the translational dynamics control loop, the aerodynamic force is weak compared with gravity and thrust, and it is difficult to obtain accurately. Therefore, the aerodynamic force is not considered in the controller design. The NDI controller of this loop can be written as

$$\dot{\mathbf{x}}_{1n} + \mathbf{x}_3 \times \mathbf{x}_1 = \frac{\mathbf{L}_{gb}^T \mathbf{G} + \mathbf{T}}{m}. \tag{10}$$

In Eq. (10), $\dot{\mathbf{x}}_{1n} = \mathbf{k}_1 \left( \mathbf{x}_1^{des} - \mathbf{x}_1 \right)$, where $\mathbf{k}_1 \in \mathbb{R}^3$ is the control gain. Define $\boldsymbol{\sigma} = \begin{bmatrix} \sigma_1 & \sigma_2 & \sigma_3 \end{bmatrix}^T$, and let

$$\boldsymbol{\sigma} = \dot{\mathbf{x}}_{1n} + \mathbf{x}_3 \times \mathbf{x}_1. \tag{11}$$

The desired control command can be obtained as

$$\begin{cases} \theta^{des} = -\arcsin\left(\sigma_1 / g\right) \\ \phi^{des} = \arcsin\left(\dfrac{\sigma_2}{\sqrt{g^2 - \sigma_1^2}}\right) \\ T_z^{des} = m\left(\sigma_3 - \sqrt{g^2 - \sigma_1^2 - \sigma_2^2}\right) \end{cases}. \tag{12}$$

The yaw motion of the coaxial rotor UAV is controlled manually; therefore, $\mathbf{x}_2^{des} = \begin{bmatrix} \phi^{des} & \theta^{des} & \psi^{ref} \end{bmatrix}^T$.

Only the effect of the rotor control moment is considered in the rotational dynamics loop, and the NDI controller can be expressed as

$$\mathbf{M}_t^{des} = \mathbf{I}\dot{\mathbf{x}}_{3n} + \mathbf{x}_3 \times \mathbf{I}\mathbf{x}_3. \tag{13}$$

In Eq. (13), $\dot{\mathbf{x}}_{3n} = \mathbf{k}_3 \left( \mathbf{x}_3^{des} - \mathbf{x}_3 \right)$, where $\mathbf{k}_3 \in \mathbb{R}^3$ is the control gain. Unlike the fixed-wing UAV, which takes aerodynamic force and moment as the primary control variables in flight, the coaxial rotor UAV takes the force and moment generated by the rotor as the direct control variables. Therefore, it is reasonable to ignore the aerodynamic force and moment, which are uncertain and difficult to calculate accurately, during the design of the control system. However, when the UAV flies fast or the natural wind speed is considerable, the aerodynamic effect is noticeable, and the NDI controller produces obvious control errors.

**3.2 INDI controller design**

Because the model's aerodynamic uncertainties and the external disturbance of UAV only exist in the dynamics loops, the controller is designed by the INDI method for the dynamics loops and the NDI method for the kinematics loops. In the translational dynamics loop, $\mathbf{x}_1$ is the state variable, $\mathbf{u}_1 = \begin{bmatrix} \theta & \phi & T_z \end{bmatrix}^{\mathrm{T}}$ is the control variable, and $\mathbf{F}$ is a function of $\mathbf{u}_1$. Taylor expansion is applied to the translational dynamic equation in the neighborhood of point $(\mathbf{x}_1, \mathbf{u}_1)$ and the first-order term is retained. We obtain

$$\dot{\mathbf{x}}_{1n} = \dot{\mathbf{x}}_1 + \left( -\frac{\partial (\mathbf{x}_3 \times \mathbf{x}_1)}{\partial \mathbf{x}_1} + \frac{1}{m}\frac{\partial \mathbf{F}}{\partial \mathbf{x}_1} \right) \Delta \mathbf{x}_1 + \left( -\frac{\partial (\mathbf{x}_3 \times \mathbf{x}_1)}{\partial \mathbf{u}_1} + \frac{1}{m}\frac{\partial \mathbf{F}}{\partial \mathbf{u}_1} \right) \Delta u_1. \tag{14}$$

According to the time scale separation method, $\Delta \mathbf{x}_1 \ll \Delta u_1$ and $\Delta \mathbf{x}_1 \approx 0$ can be obtained. Eq. (14) can be simplified as

$$\dot{\mathbf{x}}_{1n} - \dot{\mathbf{x}}_1 = \frac{1}{m}\frac{\partial \mathbf{F}}{\partial \mathbf{u}_1} \Delta \mathbf{u}_1. \tag{15}$$

Let $\dot{\mathbf{x}}_{1n} = \mathbf{k}_1 \left( \mathbf{x}_1^{des} - \mathbf{x}_1 \right)$. According to Eq. (15), the control increment can be calculated as

$$\Delta \mathbf{u}_1 = \mathbf{g}_1^{-1} \left( \mathbf{k}_1 \left( \mathbf{x}_1^{des} - \mathbf{x}_1 \right) - \dot{\mathbf{x}}_1 \right). \tag{16}$$

where

$$\mathbf{g}_1 = \begin{bmatrix} -g\cos\theta & 0 & 0 \\ -g\sin\phi\sin\theta & g\cos\phi\cos\theta & 0 \\ -g\cos\phi\sin\theta & -g\sin\phi\cos\theta & 1/m \end{bmatrix}. \tag{17}$$

According to the $\mathbf{u}_1$ in the current time and the control increment $\Delta \mathbf{u}_1$, the desired control variable of the translational dynamics loop can be written as

$$\mathbf{u}_1^{des} = \mathbf{u}_1 + \Delta \mathbf{u}_1. \tag{18}$$

By observing the above derivation, it can be found that through the process of Taylor expansion to calculate the control increment, $\partial \mathbf{A}/\partial \mathbf{u}_1 = 0$ and $\partial \mathbf{D}/\partial \mathbf{u}_1 = 0$. The aerodynamic uncertainty and the external disturbance terms are eliminated, which does not affect the calculation results of $\Delta \mathbf{u}_1$. Therefore, the INDI method has an excellent ability against aerodynamic uncertainty and external disturbance.

Similarly, in the rotational dynamics loop, the increment of the control moment can be calculated as

$$\Delta \mathbf{M}_T = \mathbf{I}\left(\mathbf{k}_3\left(\mathbf{x}_3^{des} - \mathbf{x}_3\right) - \dot{\mathbf{x}}_3\right). \tag{19}$$

The desired control moment can be written as

$$\mathbf{M}_T^{des} = \mathbf{M}_T + \Delta \mathbf{M}_T. \tag{20}$$

### 3.3 Incremental gain method

In the INDI controller shown in Eqs. (16) and (19), it is necessary to obtain the acceleration and angular acceleration of the UAV in $Ox_b y_b z_b$. $\dot{\mathbf{x}}_1$ can be directly measured by an accelerometer. $\dot{\mathbf{x}}_3$ usually needs to be obtained by the derivative of $\mathbf{x}_3$. Because there is noise in the angular velocity collected by the gyroscope, the direct derivation amplifies the noise. Therefore, a filter function needs to be added to the differentiator. The filter function causes a delay of $\dot{\mathbf{x}}_3$ and results in a severe overshoot of $\mathbf{M}_T^{des}$. The traditional method ensures the time consistency in the control loop through an active delay of $\mathbf{M}_T$, as shown in Eq. (21).

$$\begin{cases} \Delta \mathbf{M}_T^* = \mathbf{I}\left(\mathbf{k}_3\left(\mathbf{x}_3^{des} - \mathbf{x}_3\right) - \dot{\mathbf{x}}_3\left(t - \tau\right)\right) \\ \mathbf{M}_T^{des} = \mathbf{M}_T\left(t - \tau\right) + \Delta \mathbf{M}_T^* \end{cases} \tag{21}$$

In Eq. (21), $\Delta \mathbf{M}_T^*$ represents the moment increment with delay error. $\dot{\mathbf{x}}_3(t-\tau)$ represents the angular acceleration with delay $\tau$, and $\mathbf{M}_T(t-\tau)$ represents the moment with active delay $\tau$. Because $\mathbf{M}_T$ cannot be measured directly, it needs to be estimated according to Eq. (7). When the UAV aerodynamic model is not accurate, it easily causes deviations between the desired control moment and the actual control moment. In addition, when the filtering delay is unknown, the same filtering method should be applied to $\mathbf{M}_T$ to ensure the consistency of the delay, which introduces an additional computational burden to the control system.

In view of the above problems, the incremental gain method (IGM) is adopted in this paper, as shown in Eq. (22). By introducing an incremental gain $\mathbf{k}_{\Delta 3} \in \mathbb{R}^3$, the accumulated overshoot of $\Delta \mathbf{M}_T^*$ to $\mathbf{M}_T^{des}$ is alleviated, and the control error caused by the inaccurate model can be corrected.

$$\mathbf{M}_T^{des} = \mathbf{M}_T + \mathbf{k}_{\Delta 3} \Delta \mathbf{M}_T^* \tag{22}$$

# 4. Principle and characteristic of IGM

## 4.1 Discrete system stability analysis

In order to analyze the improvement principle of IGM on the system's derivative delay stability, take SISO linear system $\dot{x} = fx + gu$ as the object for analysis. The INDI controller with IGM can be expressed as

$$\begin{cases} \Delta u = g^{-1}(\dot{x}_n - \dot{x}) \\ u^{des} = u + k_\Delta \Delta u \end{cases} \tag{23}$$

In Eq. (23), $\dot{x}_n = k(x^{des} - x)$ represents the next time desired state derivative, and $x^{des}$ represents the desired state. In the practical application of the controller, $\dot{x}$ is generally obtained by central differentiation of the continuous sampling of $x$. $u$ is the actual control input of the system at the current time. Considering that the actuator system of low-cost UAVs does not have a feedback function and the actual control input cannot be measured, $u$ is equal to $u^{des}$ at the last period by default in the controller calculation. Since the INDI controller is a typical discrete control system, Z-transform is used to analyze the system's stability. The diagram of the discrete INDI control system is shown in Fig. 4.

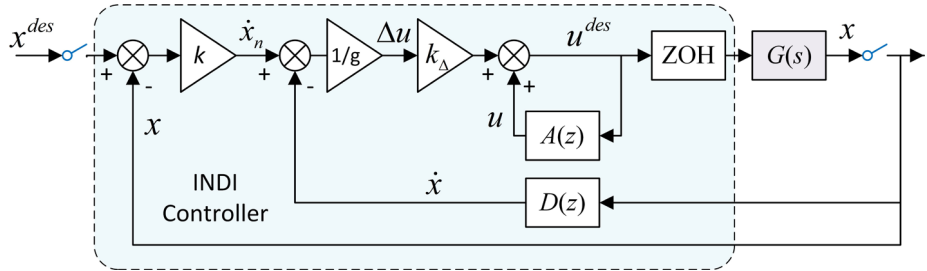

**Fig. 4 Discrete control system of INDI**

In Fig. 4, $G(s)$ represents the system's transfer function, and ZOH represents the zero-order holder. For a system with a zero-order holder, its discrete transfer function with sampling period $T$ can be written as

$$G(z) = \frac{z-1}{z} Z\left\{\frac{G(s)}{s}\right\} = \frac{z-1}{z} Z\left\{\frac{g}{s(s-f)}\right\} = \frac{-g(1-e^{fT})}{f(z-e^{fT})} \tag{24}$$

$A(z)$ represents one period delay of the input signal, which can be written as

$$A(z) = z^{-1} \tag{25}$$

$D(z)$ represents the transfer function of the differentiator. According to the central differentiator introduced

in ref [35], if the number of sampling points is $n$ ( $n = 2m+1$, $m \in \mathbb{N}^+$ ), the delay of state derivative is $\tau = mT$. This differentiator is widely used in open-source flight controllers such as Ardupilot and PX4. The INDI controller studied in this paper also adopts this differentiator. Since the more sampling points, the more complex the differentiator transfer function is, $D(z)$ is simplified, as shown in Eq. (26), to intuitively analyze the impact of state derivative delay on system stability.

$$D(z) = \frac{1}{2T}(1-z^{-2})z^{-(m-1)} \tag{26}$$

Eq. (26) represents the series connection of the three sampling points' central differentiator and delay function. It is used to simulate the delay of the differentiator as an integral multiple of the sampling period ( $\tau = mT$ ). Ref [29] regards the INDI controller as a continuous system and uses B to represent the transfer function's state derivative delay and control input delay. This paper's discrete system analysis method can reflect the relationship between the time delay and sampling frequency more accurately. According to Eq. (24) (25) (26), the closed-loop transfer function of the system shown in Fig. 4 can be written as

$$H(z) = \frac{k_\Delta k G(z)}{k_\Delta k G(z) + k_\Delta G(z) D(z) - gA(z) + g}$$
$$= \frac{-2Tk_\Delta k \left(1-e^{fT}\right) z^{m+1}}{2Tfz^{m+2} - 2T\left(k_\Delta k\left(1-e^{fT}\right) + f + fe^{fT}\right) z^{m+1} + 2Tfe^{fT}z^m - k_\Delta \left(1-e^{fT}\right) z^2 + k_\Delta \left(1-e^{fT}\right)} \tag{27}$$

As can be seen from Eq. (27), the poles and zeros of $H(z)$ are independent of $g$. Take $k_\Delta$ as the change gain, take $f = 0.5$, $k = 5$, $\tau \approx 0.1s$, and draw the root locus of the system with M = 3, 4, 5 respectively, as shown in Fig. 5.

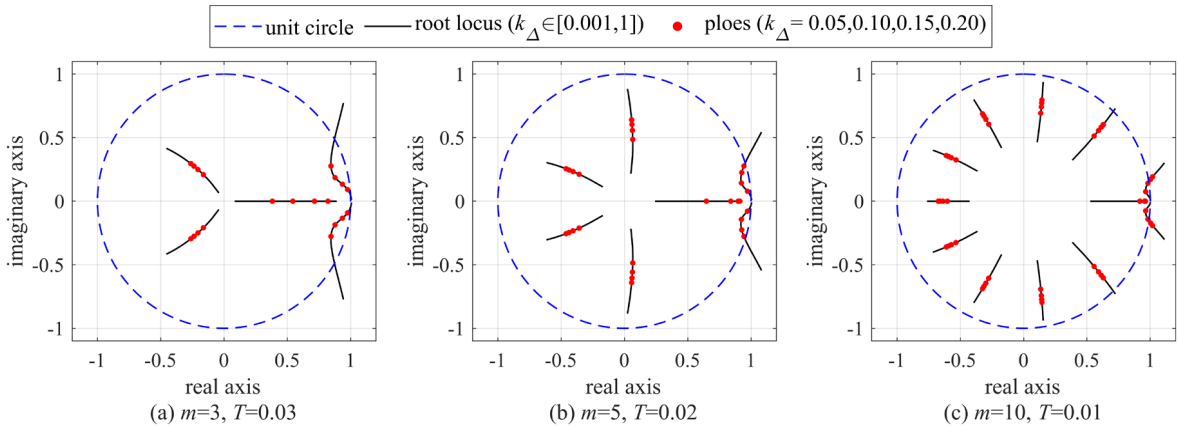

(a) $m$=3, $T$=0.03    (b) $m$=5, $T$=0.02    (c) $m$=10, $T$=0.01

**Fig. 5 Root locus of discrete INDI control system**

In Fig. 5, the solid black lines represent the root loci when $k_\Delta \in [0.001,1]$, the red dots represent the poles when $k_\Delta = 0.05, 0.10, 0.15, 0.20$, and the blue dotted lines represent the unit circle. Because $H(z)$ is the transfer function of a discrete system, the system is stable when all the system poles are located in the unit circle. As shown from Fig. 5, the number of root locus varies with different values of $m$, and the main factor causing system instability is a pair of root loci close to the right edge of the unit circle. With the change of $k_\Delta$, these two root loci first exceed the unit circle. Fig. 5 shows that for the system with state derivative delay, when $k_\Delta = 1$, the incremental gain does not work, and the system is unstable. When $k_\Delta$ takes an appropriate value, the poles can be located in the unit circle, and the system can be stable. When the total delay of state derivatives is similar, if $m = 3$, the red poles are all inside the unit circle, and if $m = 10$, the corresponding red poles of $k_\Delta = 0.15, 0.20$ appear outside the unit circle, indicating that the value range of $k_\Delta$ to make the system stable is related to $m$.

Take the system with $m = 5$ as the object, draw the root locus corresponding to different values of $T$, $f$, $k$, and the results are shown in Fig. 6.

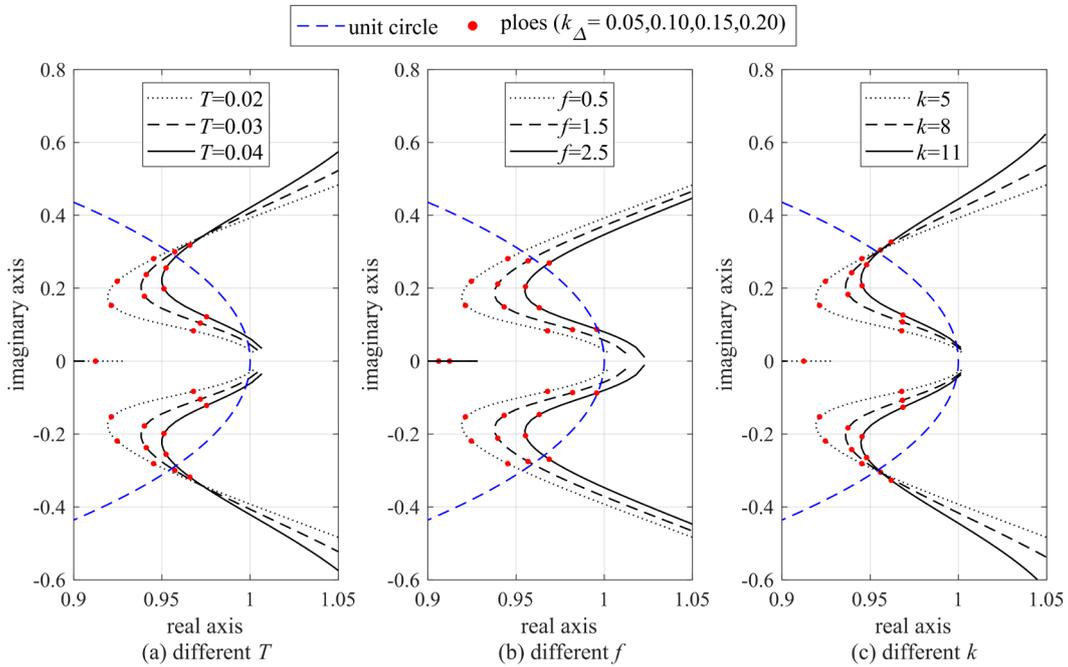

**Fig. 6 Root locus corresponding to different parameters**

The results show that different values of $T$, $f$, $k$ have a significant influence on the two root loci close to the right edge of the unit circle but have no apparent influence on other root loci. As shown in Fig. 6 (a) and (c), the smaller the values of $T$ and $k$, the more the root loci sink into the unit circle, and the more extensive $k_\Delta$ range to stabilize the system. Fig. 6 (b) shows that different $f$ values have little impact on the shape of the root loci but makes the root loci translate along the real axis. The smaller the value of $f$, the closer the root loci are to the imaginary axis, and the larger the value range of $k_\Delta$.

In order to analyze the improvement effect of IGM on system delay stability more intuitively, the influence of $k_\Delta$ on gain margin (GM) and phase margin (PM) are analyzed in the z domain. The open-loop transfer function of the system shown in Fig. 4 can be expressed as

$$H^*(z) = \frac{k_\Delta k G(z)}{k_\Delta G(z) D(z) - g A(z) + g}$$
$$= \frac{-2T k_\Delta k (1 - e^{fT}) z^{m+1}}{2Tf z^{m+2} - 2Tf(1 + e^{ft}) z^{m+1} + 2Tf e^{ft} z^m - k_\Delta (1 - e^{fT}) z^2 + k_\Delta (1 - e^{fT})} \quad (28)$$
$$= \frac{-2T k_\Delta k (1 - e^{fT}) z^{m+1}}{(z-1)\left(2Tf z^m (z - e^{ft}) - k_\Delta (1 - e^{fT})(z+1)\right)}$$

Make the system parameters $m = 5$, $T = 0.02$, $f = 0.5$, $k = 11$. The setting is exactly the same as the system parameters corresponding to the solid black line in Fig. 6 (c). Take $k_\Delta$ equals 0.10, 0.15, 0.20 respectively. The discrete system's stability analysis data is calculated and shown in Table 1. The Bode diagram and Nyquist diagram of $H^*(z)$ are shown in Fig. 7 and 8.

Table 1 stability analysis data

| $k_\Delta$ | GM (dB) | PM (deg) | Poles of open loop system | | | |
|---|---|---|---|---|---|---|
| 0.10 | 8.28 | 24.6 | 0.848+0.120i | 0.062+0.556i | -0.406+0.233i | 1.000+0.00i |
| | (14.7rad/s) | (9.37rad/s) | 0.848-0.120i | 0.062-0.556i | -0.406-0.233i | |
| 0.15 | 2.79 | 11.6 | 0.883+0.198i | 0.059+0.603i | -0.437+0.246i | 1.000+0.00i |
| | (15.0rad/s) | (13.3rad/s) | 0.883-0.198i | 0.059-0.603i | -0.437-0.246i | |
| 0.20 | -2.76 | -10.9 | 0.909+0.242i | 0.056+0.638i | -0.460+0.255i | 1.000+0.00i |
| | (15.2rad/s) | (16.5rad/s) | 0.909-0.242i | 0.056-0.638i | -0.460-0.255i | |

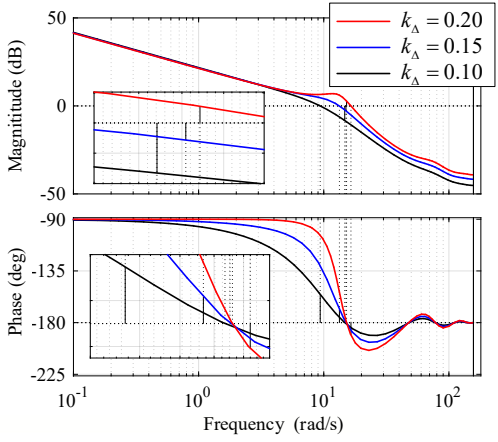 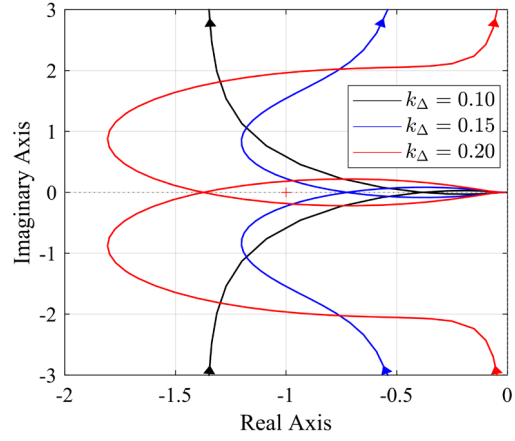

**Fig. 7 Bode diagram of discrete system**    **Fig. 8 Nyquist diagram of discrete system**

In Fig. 7, since the sampling frequency of the system is 50Hz, according to Shannon / Nyquist sampling theory, the cut-off frequency of the system is 25Hz (157rad/s). When the system is in the critical stable state, $H^*(z)$ meets $|H^*(z)|=1$ and $\angle H^*(j\omega)=-180°$. It can be seen from Fig. 7 and Table 1 that when $k_\Delta=0.10$ and $k_\Delta=0.15$, the GM and PM meet the requirements of system stability, and when $k_\Delta=0.20$, the system is unstable. This result is consistent with the result shown in Fig. 6 (c). When $k_\Delta=0.10$ and $k_\Delta=0.15$, the poles (red dots) of the closed-loop system on the root locus (solid black line) are in the unit circle, and when $k_\Delta=0.20$, the poles are outside the unit circle, indicating that the system is unstable at this time.

The system's stability can be proved more rigorously by the Nyquist diagram shown in Fig. 8. Since $H(z)$ is a discrete system, the stability condition of $H(z)$ is that when $\omega$ changes from 0 to $2\pi$, the times of $H^*(e^{j\omega})$ surround point (-1,0) in a clockwise direction must be equal to the negative number of $H^*(z)$'s poles, which are outer the unit circle [36]. According to Eq. (28), $H^*(z)$ has a pole with $z=1$, so $H^*(e^{j\omega})$ approaches infinity when $\omega=0$ and $\omega=1$. It can be seen from Fig. 8 that when $k_\Delta=0.10$ and $k_\Delta=0.15$, the number of $H^*(e^{j\omega})$ surrounds point (-1,0) is 0. According to Table 1, the number of poles outside the unit circle is 0, so the system is stable. When $k_\Delta=0.20$, if $H^*(e^{j\omega})$ is considered to be closed at infinity, the number of $H^*(e^{j\omega})$ clockwise surrounds point (-1,0) is 2, which is different from the negative value of the outer poles' number of the unit circle (0), the system is unstable.

In conclusion, the IGM can adjust the stability characteristics of the system based on the above analysis. The INDI controller with state derivative delay can be stable using $k_\Delta$ within a reasonable range.

## 4.2 Parameter selection strategy

In section 4.1, the improvement principle of IGM on the system's derivative delay stability and the influence of system parameters on $k_\Delta$ selection is proved. In practical application, the coaxial rotor UAV controller is a MIMO nonlinear system with multi feedback loops, different from SISO linear system. The difference changes the delay stability margin and the selection range of $k_\Delta$. Based on the conclusions of the previous analysis, the selection strategy of $k_\Delta$ is further analyzed from three aspects: IGM anti-delay ability, model error correction ability, and parameter adaptability in this section through the flight simulation.

In order to analyze the anti-delay ability of the IGM, the average trajectory tracking error $d_{ave}$ is defined as

$$d_{ave} = \sum_{i=1}^{n} \frac{d_i}{n}. \tag{29}$$

In Eq. (29), $d_i$ is the distance from the vehicle to the line connecting the current waypoint and the previous waypoint. $n$ is the number of samples at fixed time intervals during waypoint tracking. The size of $d_{ave}$ directly reflects the INDI controller's performance for the coaxial rotor UAV's 6-DOF motion. Since each element in $\mathbf{k}_{\Delta 3}$ has the same value, it is abbreviated as $k_{\Delta 3}$. In the waypoint tracking simulation, $\tau$ and $k_{\Delta 3}$ are taken as condition variables. The statistics result of $d_{ave}$ is shown in Fig. 9.

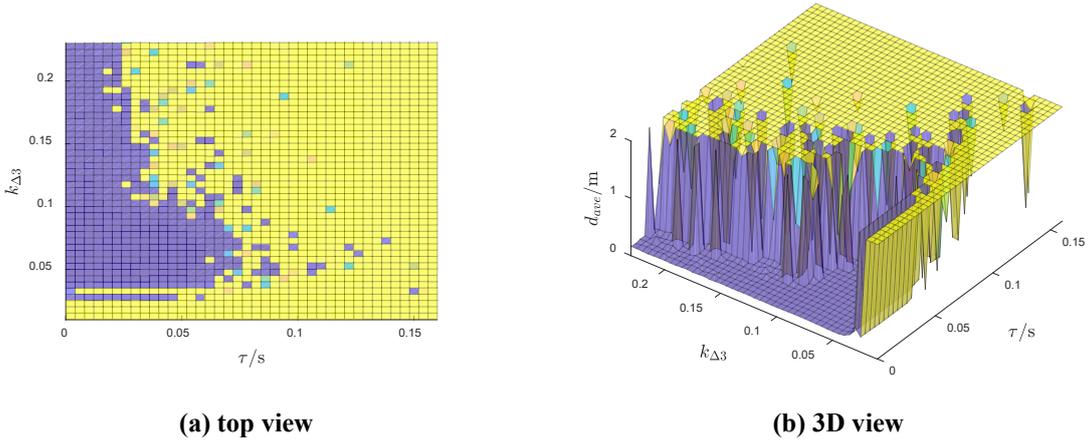

(a) top view　　　　　　　　　　(b) 3D view

**Fig. 9 Statistics result of average trajectory tracking error**

In Fig. 9, when $d_{ave} \geq 2$ (yellow area), the UAV does not have ideal waypoint tracking capability. The boundary of the purple region is defined as the time delay margin [34], and the delay on the boundary is called the maximum stable delay time ($\tau_{max}$). It can be seen from Fig. 9 that when the value of $k_{\Delta 3}$ is of the order of $10^{-2}$, the control system has ideal delay stability, and $\tau_{max}$ can reach approximately 0.06 s. A controller with a frequency of 200 Hz is used as an example. If 11 points are selected for the central differentiator [35], the derivative delay of the differentiator is 0.025 s. The IGM can fully meet the anti-delay requirements of the controller. When the value of $k_{\Delta 3}$ is of the order of $10^{-1}$, $\tau_{\max}$ decreases, which cannot meet the delay stability requirements of the controller. When the value of $k_{\Delta 3}$ is of the order of $10^{-3}$, the response of the controller slows down or fails, resulting in a sharp increase in $d_{ave}$. Based on the above analysis, the value of $k_{\Delta 3}$ is suggested to be of the order of $10^{-2}$. Nevertheless, it may not completely cover the interval (0.01, 0.1). The specific value of $k_{\Delta 3}$ is also affected by the UAV's properties and other controller parameters.

A hovering simulation is designed to analyze the model error correction ability of IGM. During simulation, $\phi^{ref} = 0$ and $\theta^{ref} = 0$. The deviation coefficient between the accurate control moment and the desired control moment of the UAV is defined as

$$\zeta = \frac{\mathbf{M}_T}{\mathbf{M}_T^{des}}. \tag{30}$$

The average angle tracking error is defined as

$$A_{ave} = \sum_{i=1}^{n} \frac{\sqrt{\phi^2 + \theta^2}}{n}. \tag{31}$$

In Eq. (31), the value of $A_{ave}$ directly reflects the control performance of INDI in the rotational dynamics loop. In the hovering simulation, $\zeta$ and $k_{\Delta 3}$ are taken as condition variables. The statistical result of $A_{ave}$ is shown in Fig. 10.

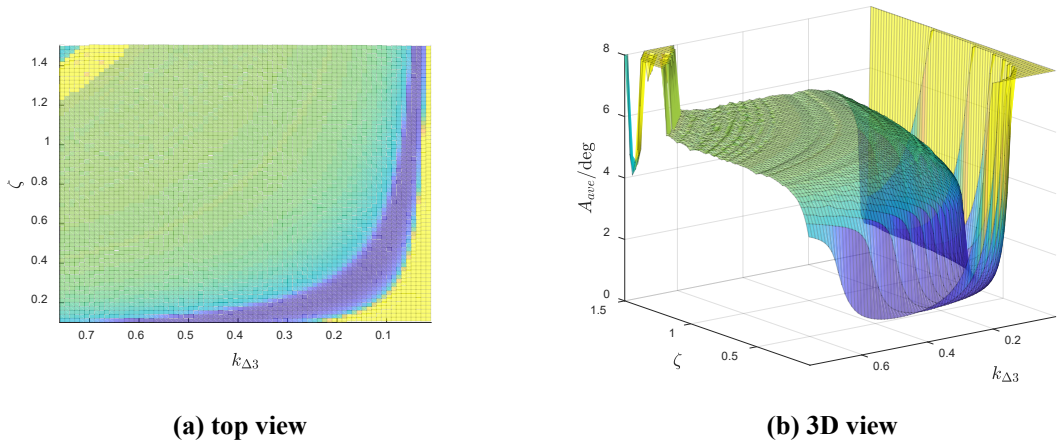

(a) top view                                  (b) 3D view

**Fig. 10 Statistics result of average angle tracking error with deviation coefficient and incremental gain**

In Fig. 10, when $A_{ave} > 5$, the UAV does not have stable hovering ability. It can be seen from the figure that when $\zeta \in [0.2, 1.4]$, there always exists a $k_{\Delta 3}$ to make $A_{ave} < 1$. When $\zeta \in [0.8, 1.4]$, the value of $k_{\Delta 3}$ that makes the vehicle hover stable, is of the order of $10^{-2}$. The above conclusions prove that IGM can correct the model error and keep the controller within ideal performance.

The hovering simulation of UAV is conducted again to analyze the parameter adaptability and convenience of IGM. Since each element in $\mathbf{k}_3$ has the same value, it is abbreviated as $k_3$. In the simulation, $k_3$ and $k_{\Delta 3}$ are taken as condition variables, and the statistical result of $A_{ave}$ is shown in Fig. 11.

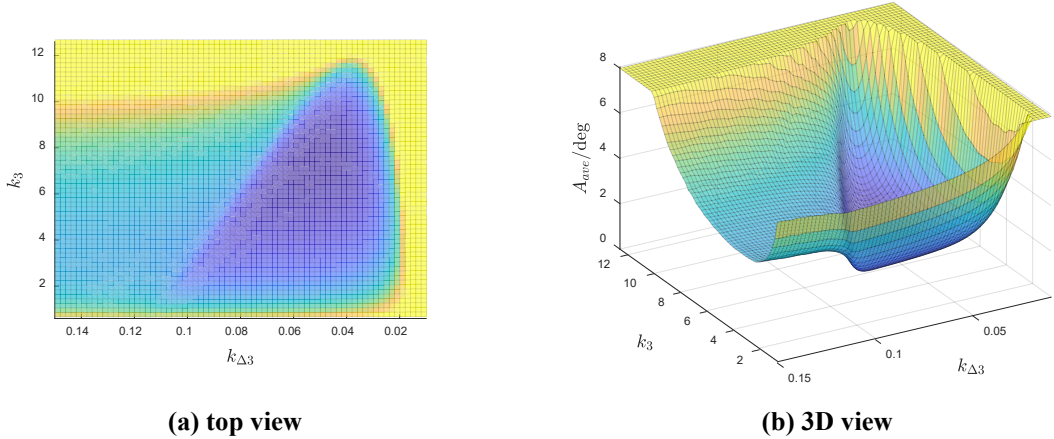

**(a) top view**          **(b) 3D view**

**Fig. 11 Statistics result of average angle tracking error with control gain and incremental gain**

It can be seen from Fig. 11 that when the value of $k_{\Delta 3}$ is of the order of $10^{-2}$ and $k_3$ is in the range of 4 to 8, the system still has good control stability, which indicates that the IGM method has good parameter adaptability and convenient parameter adjustment.

Based on the above analysis, considering the additional design constraints, the selection range of incremental gain for the stability of the MIMO nonlinear system is further reduced than that for the stability of the SISO linear system in Section 4.1. For the coaxial rotor UAV studied in this paper, the value of $k_{\Delta 3}$ is suggested to be of the order of $10^{-2}$. In the rotational dynamics loop, the IGM can keep the system stable when the angular acceleration delay is less than 0.06 s, the moment output error is in the range of - 20% to 40%, and the loop's control gain is in the range of 4 to 8.

## 5. Simulation and Flight Test

The INDI controller and NDI controller are developed based on the PX4 open-source autopilot firmware. Taking the two controllers as the control group and strictly controlling other experimental variables, and the hardware in the loop simulation, hovering flight and anti-disturbance flight tests are carried out. The control performance of the coaxial rotor UAVs and the advantages of INDI in dealing with aerodynamic uncertainty, model error, and external disturbance are verified through experiments. The IGM is used in the INDI controller to address the state derivative delay problem in the simulation and flight tests.

**5.1 Hardware in the loop simulation**

In the HITL simulation, the take-off state and external wind conditions are strictly controlled, and only the INDI and NDI are used as the control group. The random wind with the direction of $Ox_g$ extending from 3-8 m/s is generated in advance in the simulation. It is applied to INDI and NDI waypoint tracking experiments simultaneously. Under the wind effect, the aerodynamic force and moment of UAVs change unpredictably and have apparent aerodynamic uncertainty. The INDI controller's ability against aerodynamic uncertainty is verified by comparing the 6-DOF waypoint tracking performance of the two methods. The architecture of the coaxial rotor UAV HITL simulation is shown in Fig. 12.

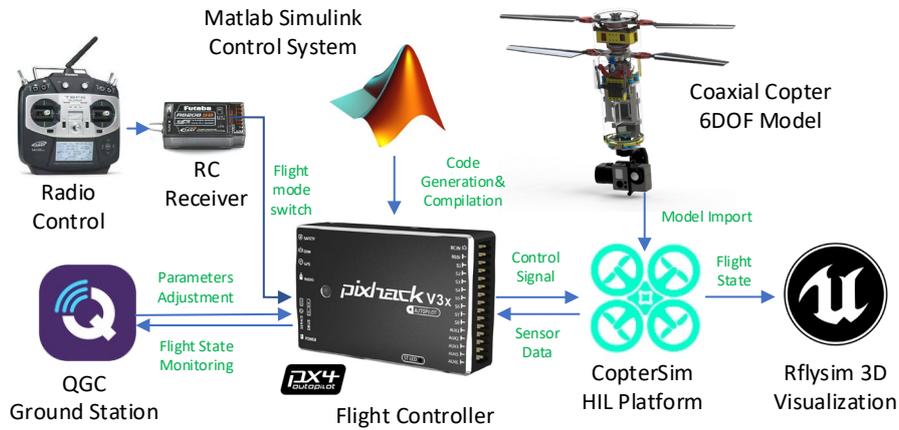

**Fig. 12 The architecture of the HITL simulation**

In the HITL simulation, the control frequency is set as 200 Hz, and the angular acceleration delay is 0.025 s. The simulation results are shown in Figs. 13-15.

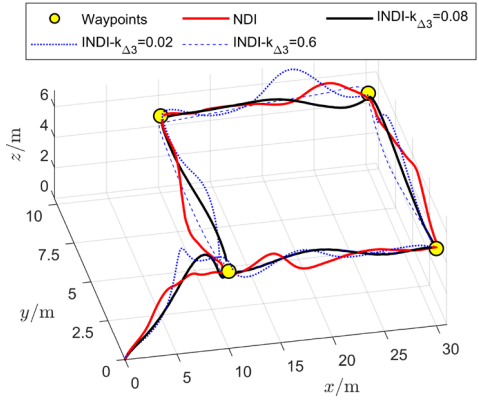
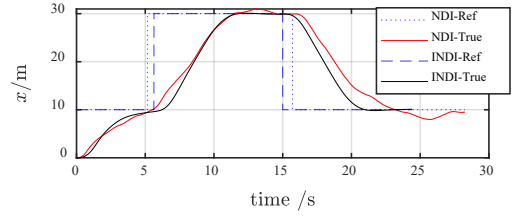

Fig. 14 Movement in ground x axis

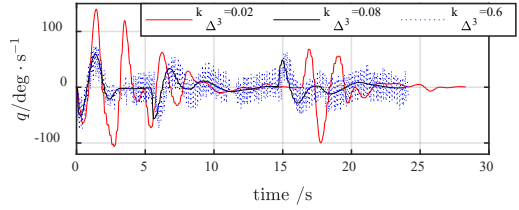

Fig. 13 Waypoint tracking trajectory

Fig. 15 Pitch angular velocity with different incremental gain

Figs. 13 and 14 show that both controllers have waypoint tracking ability. Under the same wind conditions, the trajectory of NDI has obvious fluctuations and offsets in the $Ox_g$ direction. the trajectory fluctuation of INDI is less if the appropriate gain is selected ($k_{\Delta 3} = 0.08$). The average trajectory tracking error of INDI ($d_{ave} = 0.305m$) is only 58.3% of that of NDI ($d_{ave} = 0.523m$). It can also be seen from the Figs. 13 and 15 that if the increment gain is too small ($k_{\Delta 3} = 0.02$), the control of the UAV is slow, and the fluctuations of the trajectory become larger. If the incremental gain is too large ($k_{\Delta 3} = 0.6$), the system is unstable. Although the trajectory does not diverge under the condition of the control output limitation, the fierce vibrate of angular velocity will damage the aircraft structure. The simulation results and parameter selection strategy are consistent with the analysis results in section 4.2.

**5.2 Hovering flight test**

The hovering flight test (Fig. 1) is conducted to verify the model error correction ability of the INDI controller. During the flight test, the swashplate of the UAV is intentionally set with an initial installation deviation. When the actuators are at positions 0, the swashplate does not remain horizontal. This setting makes the actual state of the UAV deviate from the control model. In the hovering flight test, only one aircraft is used, and the difference in natural wind is ignored in each flight. The comparative flight tests of INDI and NDI are carried out by replacing the control program.

The results of the NDI hovering flight test are shown in Fig. 16.

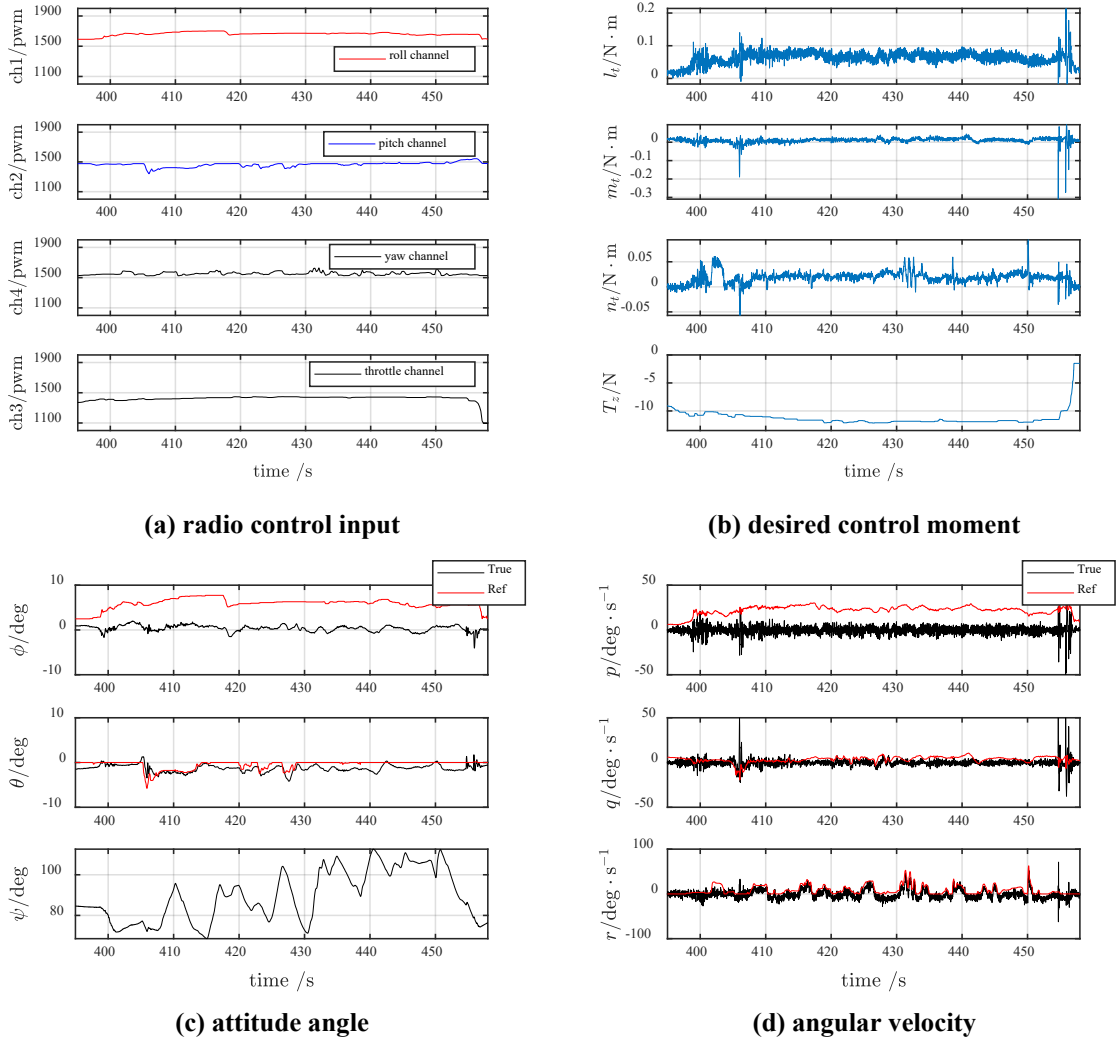

**(a) radio control input**  **(b) desired control moment**

**(c) attitude angle**  **(d) angular velocity**

**Fig. 16 results of NDI hovering flight test**

In the hovering flight test, $\phi^{ref}, \theta^{ref}, r^{ref}, T_z^{ref}$ are controlled manually. Fig. 16 (c) shows that an obvious steady-state error exists in the roll channel of the NDI controller. The reason is that NDI depends on the model accuracy and cannot automatically correct the model error. The pilot needs to manually add an offset in the desired control command to keep the UAV hovering stably. In Fig. 16 (a), each channel's PWM input ranges from 1100 to 1900, and the roll channel's median PWM is 1500. The pilot uses nearly 25% of the input signal to correct the model error during flight tests.

The results of the INDI hovering flight test are shown in Fig. 17.

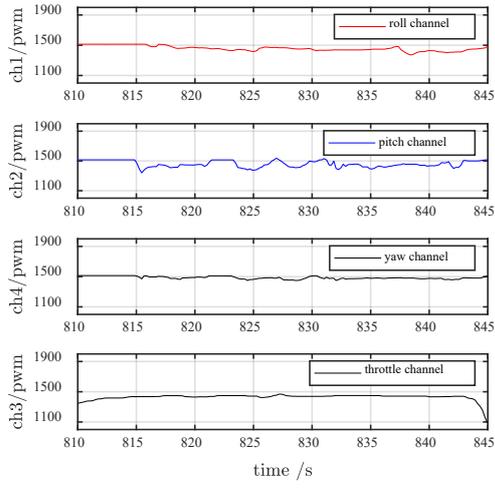

**(a) radio control input**

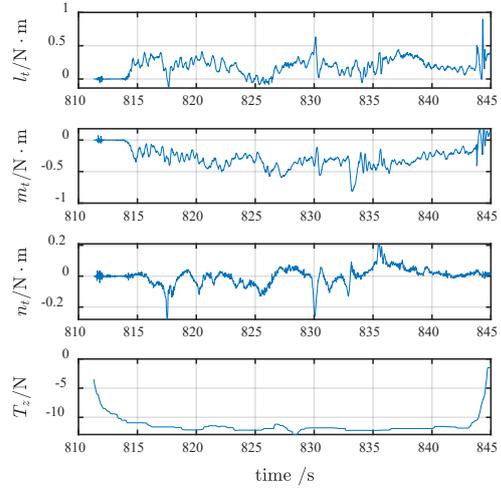

**(b) desired control moment**

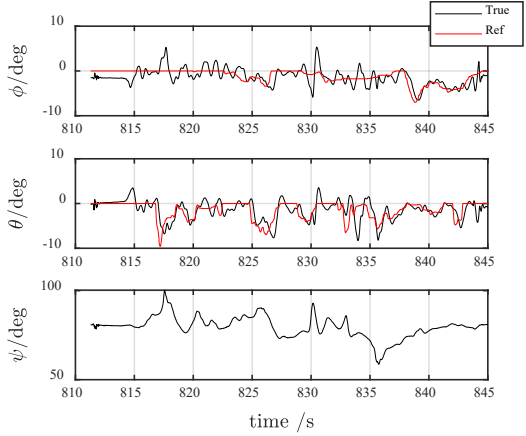

**(c) attitude angle**

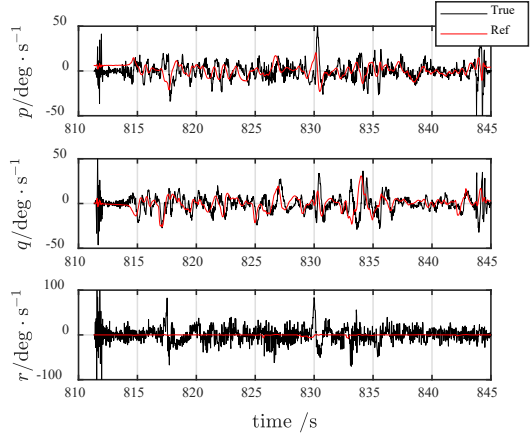

**(d) angular velocity**

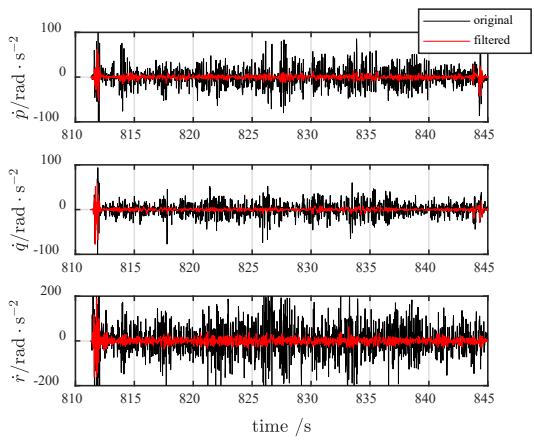

**(e) angular acceleration**

**Fig. 17 Results of INDI hovering flight test**

Figs. 17 (c) and (d) show that the UAV has good tracking ability for the desired attitude angle and the desired angular velocity. There is no steady-state error, which indicates that the INDI has excellent model error correction ability. In an actual flight, the central differentiator is used to calculate the angular acceleration. In the differentiator, 11 sampling points with an interval of 0.004 s are adopted. The delay of the differentiator is 0.02 s. The calculation result of angular acceleration is shown in Fig. 17 (e). The black line in the figure represents the direct derivative result, and the red line represents the result of the differentiator. The filtering function of the differentiator greatly reduces the noise of the angular acceleration. During the flight test, IGM is used to overcome angular acceleration delay, giving the INDI controller a good control quality. Furthermore, the control parameter selection strategy is consistent with the analysis results in section 4.2.

## 5.3 Anti-disturbance flight test

In the anti-disturbance flight test, $\phi^{ref}=0$, $\theta^{ref}=0$, $r^{ref}=0$ is set, and the pilot inputs the disturbance command through radio control. Moreover, the flight experiment is conducted under windless indoor conditions to ensure that the radio control command is the only disturbance source. In the case that the controller is completely unknown to the strong external disturbance, the anti-disturbance performance of the INDI controller is verified. The input principle of the remote disturbance command is shown in Fig. 18, and the environment of the anti-disturbance flight test is shown in Fig. 19.

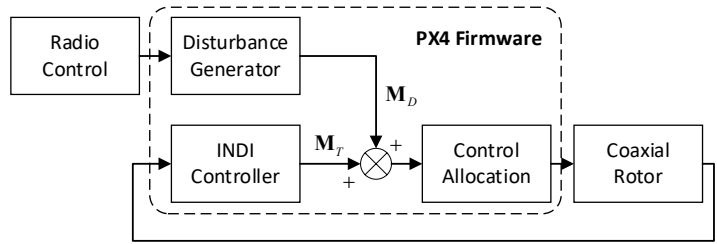

**Fig. 18 The input principle of the remote disturbance command**

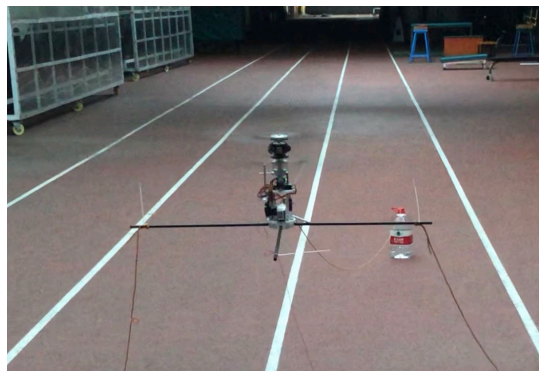

**Fig. 19 INDI anti-disturbance flight test**

The results of the anti-disturbance flight test are shown in Fig. 20.

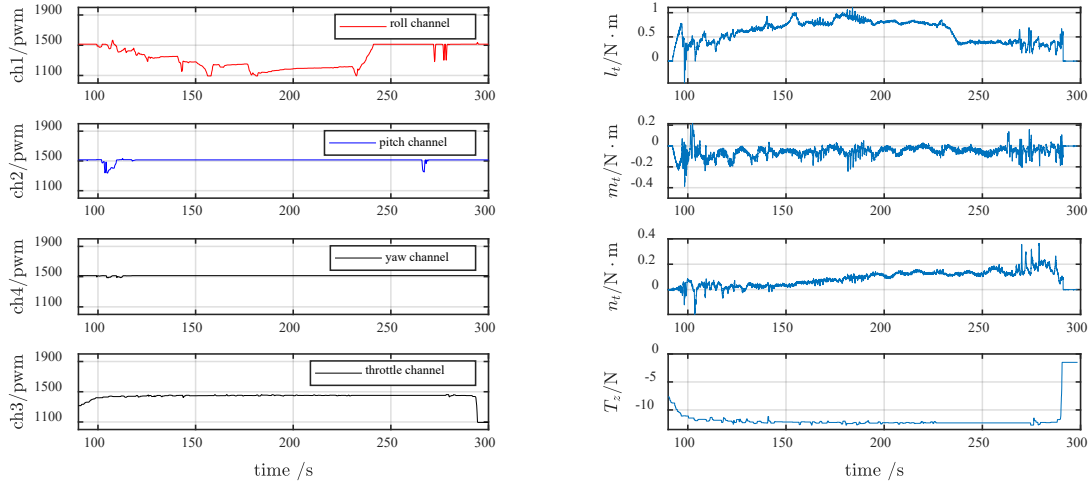

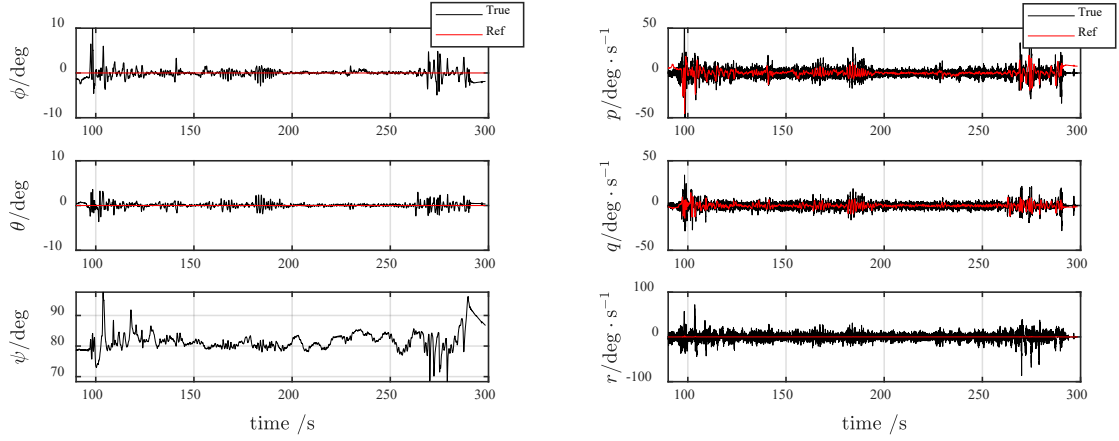

**(c) attitude angle**    **(d) angular velocity**
**Fig. 20 Results of INDI anti-disturbance flight test**

Figs. 20 (a) shows that an obvious disturbance command is imposed on the roll channel during the flight test. Fig. 20 (b) shows that the INDI controller can use the accumulation of the control increment to adjust the output quickly to overcome the influence of the disturbance. In the anti-disturbance flight test, the UAV maintains good flight stability; the attitude angle and angular velocity only fluctuate slightly (Figs. 20 (c) (d)). The flight test proves that the INDI controller has excellent anti-disturbance ability.

## 6. Conclusion

In this paper, INDI and NDI controllers are designed for a coaxial rotor UAV. The IGM method is proposed to solve the delay problem of the state derivative in the INDI controller, and the incremental gain selection strategy is analyzed. The advantages of INDI in the application of coaxial rotor UAVs are verified through comparative experiments of the two controllers. The conclusions are as follows.

(1) The INDI and NDI controllers designed in this paper can make the coaxial rotor UAV have ideal waypoint tracking and hovering flight ability.

(2) The value of the elements in the incremental gain $\mathbf{k}_{\Delta 3}$ is recommended to be of the order of $10^{-2}$. The flight state vibrates If the value of $\mathbf{k}_{\Delta 3}$ is too large, while the tracking performance of the controller becomes worse if the value is too small.

(3) In the rotational dynamics loop, the incremental gain method can keep the system stable when the angular acceleration delay is less than 0.06 s. the moment output error is in the range of -20% to 40%, and the loop's control gain is in the range of 4 to 8.

(4) Compared with the NDI, the INDI has better abilities to overcome aerodynamic uncertainty, correct for model errors, and handle disturbance.

This work proves the principle of IGM in the SISO linear system and verifies its application in MIMO nonlinear system. In future work, the stability of the INDI controller with IGM should be further analyzed in MIMO nonlinear system, and more accurate incremental gain selection results should be given.

## Acknowledgments

This work is supported by the National Natural Science Foundation of China under Grant Nos. 11872230, 91852108 and 92052203.